\documentclass[12pt]{iopart}
\usepackage{graphicx}
\usepackage[T1]{url}
\usepackage{subfigure}
\usepackage{multibib}

\newcommand{\be}{\begin{equation}}
\newcommand{\ee}{\end{equation}}
\newcommand{\beu}{\begin{displaymath}}
\newcommand{\eeu}{\end{displaymath}}

\newcommand{\beq}{\begin{eqnarray}}
\newcommand{\eeq}{\end{eqnarray}}

\newcommand{\equref}[1]{equation \ref{eq:#1}}
\newcommand{\figref}[1]{figure \ref{fig:#1}}
\newcommand{\tabref}[1]{table \ref{tab:#1}}
\newcommand{\secref}[1]{section \ref{sec:#1}}
\begin{document}
\bibliographystyle{unsrt}
\title[Bond analysis of cobalt and iron based skutterudites]{Bond
  analysis of cobalt and iron based skutterudites: elongated lanthanum
  bonds in LaFe$_4$P$_{12}$}
\author{Espen Flage-Larsen, Ole Martin L{\o}vvik, {\O}ystein Prytz and
Johan Taft{\o}}
\address{Department of Physics, University of Oslo, P O Box 1048
  Blindern, NO-0316 Oslo, Norway}
\ead{espen.flage-larsen@fys.uio.no}
\date{\today}
\begin{abstract}
Motivated by the possibility of further improving the thermoelectric properties of skutterudites
we investigate charge transfer and bonding in this class of materials using density functional calculations. Results for the CoP$_3$, CoSb$_3$, LaFe$_4$P$_{12}$ and the hypothetical FeP$_3$ compounds are presented
using the procrystal as the non-binding reference. Spherical integration and Bader analysis are
performed to illustrate charge transfer differences between these compounds. The results are in
good qualitative agreement with simple electronegativity considerations. The calculations confirm
that the transition metal-pnictogen and the pnictogen-pnictogen bonds are covalent, while the filler
atom-pnictogen bond is of a more polar and complex nature. The success
of the ``rattling'' cage as phonon inhibitor
is explained by a unique semi-correlated bonding scheme
between lanthanum and phosphorus. Elongated bonds along the crystal axes
surrounds the lanthanum ion and generate a dodecahedra
grid. Vibrations along the crystal axes are then closely connected to
and scatter from the phosphorus rings. In the other directions, a more uncorrelated
vibration is possible. This duality widens the
phonon dampening possibilities.
\end{abstract}
\submitto{\NJP}
\maketitle
\section{Introduction} 
In recent years, skutterudite compounds have attracted much attention as potential thermoelectric
materials\cite{exp:snydercomplex}.
This is mainly attributed to their high
figure-of-merit\cite{exp:snydercomplex} and the possibility to substitute and
modify the composition while still retaining the basic
structure\cite{theo:olemfilled}. The figure-of-merit is defined as $ZT = \alpha^2 \sigma T/\kappa$, where $\alpha$, $\sigma$, $\kappa$ and $T$ are the Seebeck coefficient,
the electrical conductivity, the thermal conductivity, and
the temperature, respectively. The high $ZT$ value comes as a result of a strong asymmetric density of
states around the Fermi energy and the ability to drastically decrease the thermal conductivity by adding interstitial atoms. These structures are also interesting from
a physical point of view due to superconducting\cite{exp:shirotaniskutt,exp:meisner},
heavy-fermion\cite{exp:gajewski,exp:morelli} and ferromagnetic\cite{exp:danebrock} behaviour.

Empty skutterudite compounds in space group $Im\overline{3}$ have the chemical formula $MX_3$, where $M$
and $X$ are transition metals and pnictogens, respectively.

\begin{figure}[ht]
\centering
\includegraphics[width=0.8\textwidth]{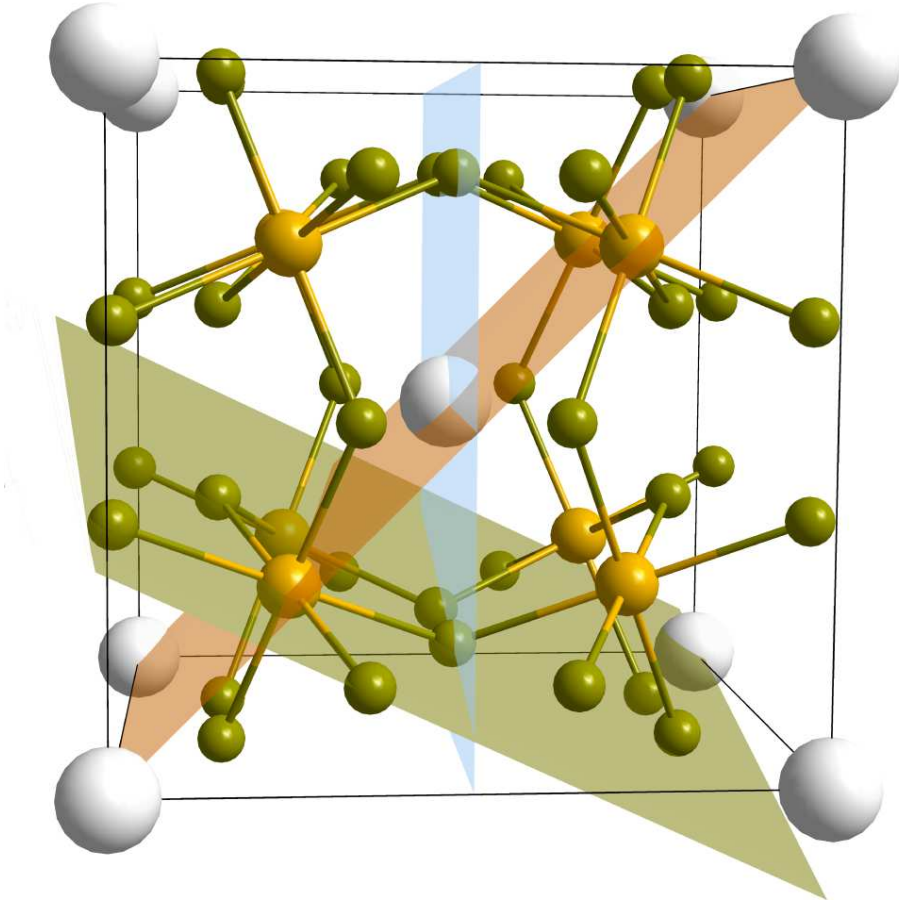}
\caption{(Color online) The skutterudite structure. Small,
medium and large spheres are the pnictogens, transition metals and filler
ions, respectively. The octahedral arrangement of pnictogens
around the transition metal is shown in addition to the octahedral cutout (green/darkest), the 020 (blue/brightest) and the 110
(red/bright) planes.}
\label{fig:skutt}
\end{figure}

It was early suggested\cite{exp:slackfiller} that the inclusion of heavy
atoms in the open voids, the positions of the filler ions
in \figref{skutt}, could reduce the thermal conductivity. This
idea has since been confirmed by numerous studies\cite{exp:uher}.  The
decrease of the thermal conductivity is attributed to a ``rattling''
behaviour of the filler atoms, confirmed by inelastic neutron
diffraction studies\cite{exp:keppensratlers}.

The true nature of the rattling
behaviour has not yet been determined, but it has been suggested that the filler ions can rattle
independently\cite{exp:saleskeppens} in the dodecahedra of
$X$ ions. Recent
studies\cite{exp:koza,exp:chris} contradict this picture
and suggest a more correlated motion of the filler atoms. 

Changes of the thermal conductivity have also been shown for substitutions in the basic skutterudite
framework\cite{exp:stiewesubst,exp:christensensubst}. Such substitutions can also be tuned
to modify the electrical conductivity\cite{exp:annoconduct} by influencing
the electronic part of the thermal conductivity. Hence
both the electrical and thermal conductivity can be tailored by altering the composition.
The wealth of different compositions possible in the
skutterudite structure has resulted in a number of studies looking for optimal compositions to reach ultimate
thermoelectric properties\cite{exp:sales,exp:uher}. To assist in the
search, 
an in-depth understanding of the electron structure and
lattice thermal conductivity may be helpful. 

In this work
we focus on the valence electron density and bonding
properties. Previous work has indicated the importance
of the hybridization between the p- and d-orbitals of the
antimonides and transition metals to stabilize the CoSb$_6$
octahedra\cite{theo:devos}. The lowest conduction band responsible
for electron transport was furthermore shown to include
mainly Sb p-states located in the Sb$_4$ ring\cite{theo:devos}. These
rings where also shown to be connected through the
CoSb$_6$ octahedra by the highest valence band. In addition, evidence of delocalized non-bonding Sb s-states was
observed. 

There has been some discussions\cite{theo:llunell} as to whether
it is the $MX_6$ octahedra, the $X_4$ ring or a
combination of both\cite{theo:devos} that determine the main bonding features of the skutterudite compounds. Numerous
studies\cite{exp:grosvenor,exp:oysteincompexp} have calculated and measured
the density of states around the Fermi level. From these
studies it is quite clear that the 3d-states of $M$ lie right
under the Fermi level followed by the 3p-states of $X$,
while the $X$ 3s-states lie deep below the Fermi level and
are not contributing to any significant bonds. The lower
part of the conduction band has primarily 3d- and 3p-state character from $M$ and $X$ respectively. 

However, to our knowledge the bonding character of the filler ions has not been specifically addressed in detail beyond
what has been suggested in literature reviews\cite{exp:uher,
  exp:nolas}, in spite of the filler atoms' special role as enhancing
lattice thermal resistivity. Recent studies\cite{exp:koza,exp:chris}
suggest a more correlated movement of the filler atoms. This motivates us to specifically address the
filler ion charge transfers. 

In this study we investigate the electronic structure
and charge transfer in the CoP$_{3}$, CoSb$_{3}$, LaFe$_{4}$P$_{12}$
and the hypothetical FeP$_{3}$ skutterudites. We
give an introduction to the DFT and charge transfer calculations in
\secref{dft} and \secref{chgana}, respectively. Results
and discussions follow in \secref{results} where we briefly list the
relaxed structure in \secref{relaxation} followed by a charge transfer reference discussion in \secref{transferref}. Results of a qualitative charge transfer analysis using the procrystal as a
reference are presented in \secref{chgtransfer} followed by a more
quantitative analysis in \secref{quantitative}, which cover both
estimates of charge depletion numbers and Bader analyses. Brief
discussions finally ends \secref{results}. Concluding remarks are given in \secref{conclude}
\section{Computational details}\label{sec:compdetails}
\subsection{Density functional theory}\label{sec:dft}
The electronic structure and bonding information are
ultimately manifested in the charge density, the fundamental quantity in density functional theory (DFT)\cite{dft:ks,dft:kohn}. This makes DFT an obvious choice for charge density analysis. We have performed DFT calculations in the
generalized gradient approximation (GGA)\cite{dft:gga2,dft:gga4,dft:gga3,dft:gga1} using
Perdew-Burke-Ernzerhof (PBE)\cite{dft:pbe} exchange-correlation
functional to obtain the charge density. The highly efficient projector-augmented-wave (PAW)\cite{dft:paw,vasp:vasp5} method
was also used. All calculation were done using the Vienna
Ab initio Simulation Package
(VASP)\cite{vasp:vasp5,vasp:vasp3,vasp:vasp4,vasp:vasp2,vasp:vasp1,vasp:homepage}. 

Crystal
structures were relaxed with regards to cell size, shape
and atomic positions using the residual minimization
scheme, direct inversion in the iterative subspace (RMM-DIIS)\cite{vasp:rmmdiis} algorithm. After the relaxation another self-consistent calculation has been done to ensure correct
representation of the system. An energy cutoff of 800
eV for LaFe$_4$P$_{12}$ and 550 eV for CoP$_3$ and CoSb$_3$ were
necessary to obtain convergence of the total energy to within 2 meV. A k-point grid of 8$\times$8$\times$8 for the conventional (cubic) unit cell was sufficient to achieve similar
convergence.

Comparable smearing conditions are very important
during charge analysis, since different smearing parameters will affect the charge density well above errors introduced in the post-processing. A Gaussian smearing\cite{vasp:gaussian} of
0.3 eV was sufficient to converge all structures and atoms
to 1 $\mu$eV within fifty electron steps. The difference between Gaussian
and higher order Methfessel and Paxton smearing\cite{vasp:mp} was found to be very small for LaFe$_4$P$_{12}$ ,
thus Gaussian smearing was used in all cases. Procrystal
electron densities for charge transfer analysis were generated from free atom densities obtained from large single
atom unit cell calculations using VASP. The lattice constants were scaled two times that of the crystal to avoid
any bonding with mirror images. To make sure that the
resolution was comparable for the augmentation charges,
the grid size was adjusted accordingly.
\subsection{Charge transfer analysis}\label{sec:chgana}
The charge transfer $\rho_{b}$ in a material can be defined by
\be \label{eq:chargebond}
\rho_{b}=\rho_{0}-\rho_{r},
\ee 
where $\rho_r$ is a reference density and $\rho_{0}$ is the crystal
charge density of the material. Bonding properties can then be
quantified and visualized from $\rho_b$. The choice of $\rho_r$ is
important and is discussed in \secref{transferref} 

The charge density in three dimensions is difficult to
present without the use of proper visualization tools. In this work the charge density is represented by interionic line extractions of the charge density using modified
Shepard interpolation\cite{num:renka} or by a scalar contour plot in
the planes defined in \figref{skutt}. The interpolation routine is
used to force correct extraction exactly along the line.

The problem of assigning quantitative charge to an ion
and thus determine electron transfer is well known, and
a number of different approaches are being used\cite{theo:carlonewdir}. In
this work we will use a Bader analysis\cite{theo:bader} scheme to determine the quantitative ionic electron occupancy numbers. The underlying principle of the Bader analysis is
based on a division of charge surfaces where the gradient
of the charge density does not have a component normal
to its surface. Recently Bader analysis has been implemented efficiently\cite{theo:baderprog3,theo:baderprog2,theo:baderprog1} in a freely available code\cite{theo:baderhomepage}.
The Bader analysis can sometimes be difficult to interpret and a correct representation of the core charges is
necessary.

As an alternative, to determine the charge depletion it
is possible to simply integrate the charge density around
a single ion inside a sphere. This raises two fundamental objections. In general the bonds are not spherically
symmetric. Also, the unit cell volume can not be covered by spheres alone, which means that parts of the
electron density must escape the analysis. Nevertheless
a spherical body shape is simple to define and intuitively
easy to analyse. In this work we have therefore used the
spherical integration technique as an alternative to calculate the
depletion around each ion, well aware of its limitations. For spherical calculations the radial cutoffs are crucial to obtain representative depletion
numbers. A common approach is to use the covalent radius as a the cutoff radius. This is obviously a simplification and may be unreasonable for bonds not obeying
covalency. 

In this work we suggest an alternative way to
determine the radial-cutoff $r_d$. We start by integrating
the charge density difference $\rho_b$ from which information
about the charge transfer is directly available. We define
the integral $I_b$ as
\be
\label{eq:chargeint}
I_{b}(r_{max})=\int_{\Omega} \int_{0}^{r_{max}} \rho_{b}drd\Omega,
\ee
where the volume of integration is limited by the outer radius cutoff
$r_{max}$. The angle dependence is handled by $d\Omega$.
The electron charge depletion number $\Delta n_{d}$ is now defined as follows
\be \label{eq:diffchange}
\Delta n_{d}=I_{b}(r_{max}=r_d).
\ee
In this work $r_{d}$ is defined as the radius where the integral $I_b$ peaks. More specifically as 
\be \label{eq:occexpectation}
\left.\frac{\partial I_{b}}{\partial r_{max}}\right|_{r_{max}=r_{d}}=0.
\ee
We require that $r_d > r_t$, a lower threshold limit. Choosing this threshold limit can be difficult in general. We
have found that a value of $r_t$ = 0.3 {\AA} avoids possible local oscillations close to the core giving rise to local extremal
values. Controllable accuracy during the integration has
been obtained by the use of a numerical trapezoidal integration scheme. This relies on the modified Shepard
interpolation routine which is able to return any given
function value after the initialization of the basis functions, thus allowing a correct representation close to the
integration limits without severe performance penalties.
Furthermore, we have calculated the mean value of the
depletion $\langle \Delta n_d\rangle$ by integrating $I_b$ such that
\be
\label{eq:dipoleeq}
\langle \Delta n_{d}\rangle=\frac{1}{r_{d}}\int_{0}^{r_{d}}I_{b}(r_{max}) dr_{max},
\ee

Typically a small difference between $\langle \Delta n_b \rangle$ and
$\Delta n_d$ indicates an abrupt and short-ranged depletion zone
around the ion (the
opposite is true for a larger difference). This and the value of $r_d$ give information about the extension
of the depletion. An obvious limitation of this method is
that occupancy numbers can not be obtained.
To compare charge transfer maxima locations and their
covalent character we have defined the center of Pauling
electronegativity as $\chi^* = \chi_1 /(\chi_1 + \chi_2)$, where $\chi_1$ and $\chi_2$
are the electronegativities of the respective ions.

The need for a correct representation of the all-electron
charge density must be emphasized. Due to the compensator charge density\cite{vasp:vasp5} the usual charge density obtained
from VASP is not the true all-electron charge density. In
this work the all-electron charge density has been explicitly regenerated after a pre-converged run, thus removing the problems associated with the compensator charge
density. For the Bader analysis the total charge density (valence+core) have been included as a reference to
ensure proper determination of the maximum and minimum valence electron density. For all other calculations
in this work the separated all-electron valence charge density has been used exclusively.

It should be mentioned that other geometries are easily
adopted in this method and work is in the progress to
extend the analysis to ionic basins\cite{theo:badercarlo} and other body
shapes, still using a selected reference charge density.
\section{Results and discussion}\label{sec:results}
\subsection{Structural relaxation}\label{sec:relaxation}
Experimental lattice parameters and atom positions
were used as a start for the structural relaxation. Results were in good agreement with previous experimental
measurements\cite{exp:schmidt,exp:jeitschko2,exp:jeitschko} and calculations\cite{exp:oysteincompexp}; lattice constants and Wyckoff positions of all relaxed structures
were within one percent of the experimental value. In
agreement with earlier reports\cite{exp:schmidt,exp:jeitschko2,exp:jeitschko}, the P$_4$ ring becomes more quadratic going from CoP$_3$ to LaFe$_4$P$_{12}$. For
these materials the width to length ratio (ring ratio) is 0.79 and 0.97,
respectively. The Sb$_4$ ring ratio is 0.96 for CoSb$_3$, in agreement
with previous experimental work\cite{exp:ohno}. 

Increased deviation from quadratic rings between, CoP$_3$ and CoSb$_3$ are expected due to the larger
atoms dominating influence on the short bond. This increases the two
shortest bond lengths, yielding a more quadratic shaped ring.
\subsection{Choice of charge transfer reference}\label{sec:transferref}
The determination of charge transfer is a difficult process due to the 
choice of charge reference. From a theoretical point of
view we can use a procrystal\cite{exp:johanvalence} charge density
$\rho_p$. This is generated from a superposition of free atomic charge densities embedded in the crystal unit cell, similar to the
independent atom model\cite{exp:iam}. These charge densities will
overlap, but bonding features are absent. This is intuitive and generality is preserved. The bonding properties
can be determined from \equref{chargebond} using $\rho_p$ as the reference
charge density $\rho_r$.

Different experimental methods exist to determine the
charge density and/or charge transfer. One approach is
to compare spectroscopic data with standard state references $\rho_m$\cite{exp:grosvenor,exp:annoxray,exp:graetz,exp:oystein}. The relative intensity difference may then be converted to occupancy numbers\cite{exp:pearson}.
A more direct approach is to perform diffraction experiments and refine the structure factors by starting from
overlapping atomic orbitals\cite{exp:johanvalence,exp:zuocharge,exp:zuostruct}. This reference
is in principle equivalent to the procrystal used in this
work. The structure-factor refinement is stopped when
sufficient agreement with experimental diffraction intensities is
reached. A standard state charge density $\rho_m$ reference will
already contain bonding features on the reference level.

\begin{figure}[ht]
\centering
\includegraphics[width=0.8\textwidth]{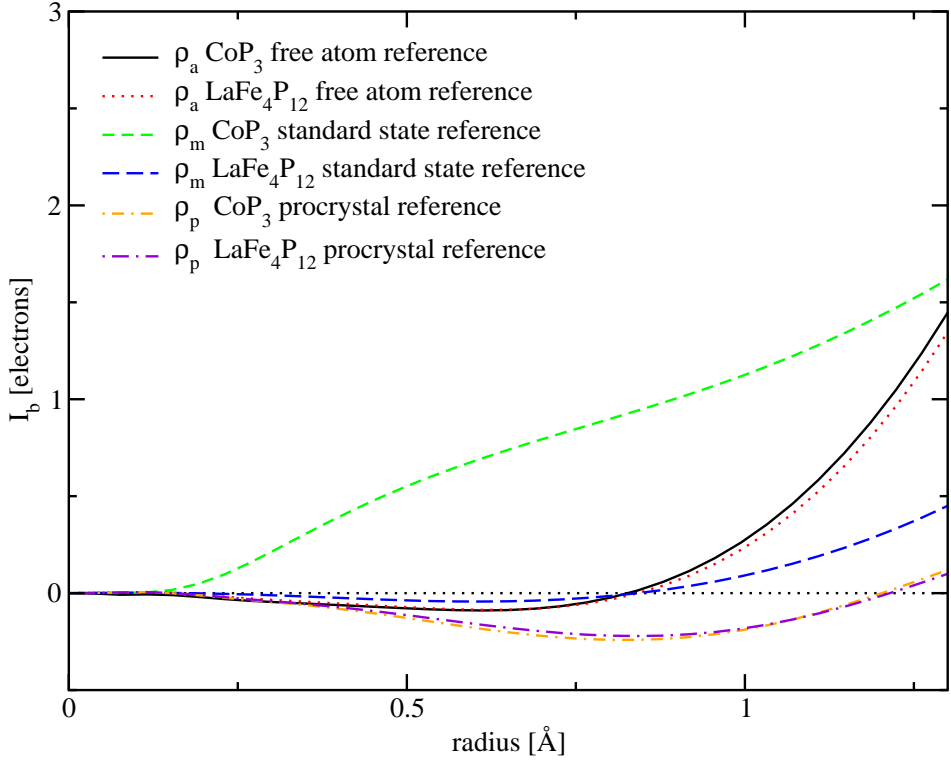}
\caption{(Color online) The effect of different charge references
  $\rho_r$ on the integrated charge density difference $I_b$ around
  the transition metal. References employed are the standard
state $\rho_m$ (long-dashed line for LaFe$_4$P$_{12}$, short-dashed line
for CoP$_3$), the atomic $\rho_{a}$ (dotted line for LaFe$_4$P$_{12}$, solid line
for CoP$_3$) and the procrystal $\rho_p$ (long-dash-dotted line for
LaFe$_4$P$_{12}$, short-dash-dotted line for CoP$_3$).
}
\label{fig:spher}
\end{figure}

Comparisons between different compounds are then complicated by the lack of a common bond-free reference. To
illustrate the complications we have calculated the spherically integrated charge density difference $\rho_b$ around a $M$
ion for LaFe$_4$P$_{12}$ and CoP$_3$ using the free atom $\rho_{a}$, the
standard state $\rho_m$ and the procrystal $\rho_p$ charge
density references. Moving from the $\rho_p$ reference to $\rho_m$
a decrease depletion around the $M$ ion is expected due to the
underlying bonding features in $\rho_m$. The charge around an atom in
the procrystal can be anticipated to be larger than
around the free atom due to the overlapping free atom charge densities. Both are clearly confirmed in \figref{spher}. In the case of
LaFe$_4$P$_{12}$ a charge depletion around Fe is observed using
both the standard state Fe and the procrystal as references. However, the magnitude and gradient are different between the two references. More serious problems
emerge when standard state Co for CoP$_3$ is
used. A charge buildup close to Co ion is then found, opposite to what is seen from the procrystal results. The
two references thus represent two different pictures; a
buildup and a depletion of charge respectively.

Hence, if we intend to compare charge transfer magnitudes between CoP$_3$ and LaFe$_4$P$_{12}$ we should preferably
use $\rho_p$ or $\rho_{a}$ as a reference. The procrystal represents a consistent reference in charge transfer
analysis, but may be difficult to use directly as a reference experimentally. All further charge transfer analysis
in this work use the procrystal $\rho_p$ as the reference charge
density $\rho_r$.
\subsection{Qualitative charge transfer analysis}\label{sec:chgtransfer}
\begin{figure}[ht]
\centering
\subfigure[]{\fbox{\includegraphics[width=0.65\textwidth]{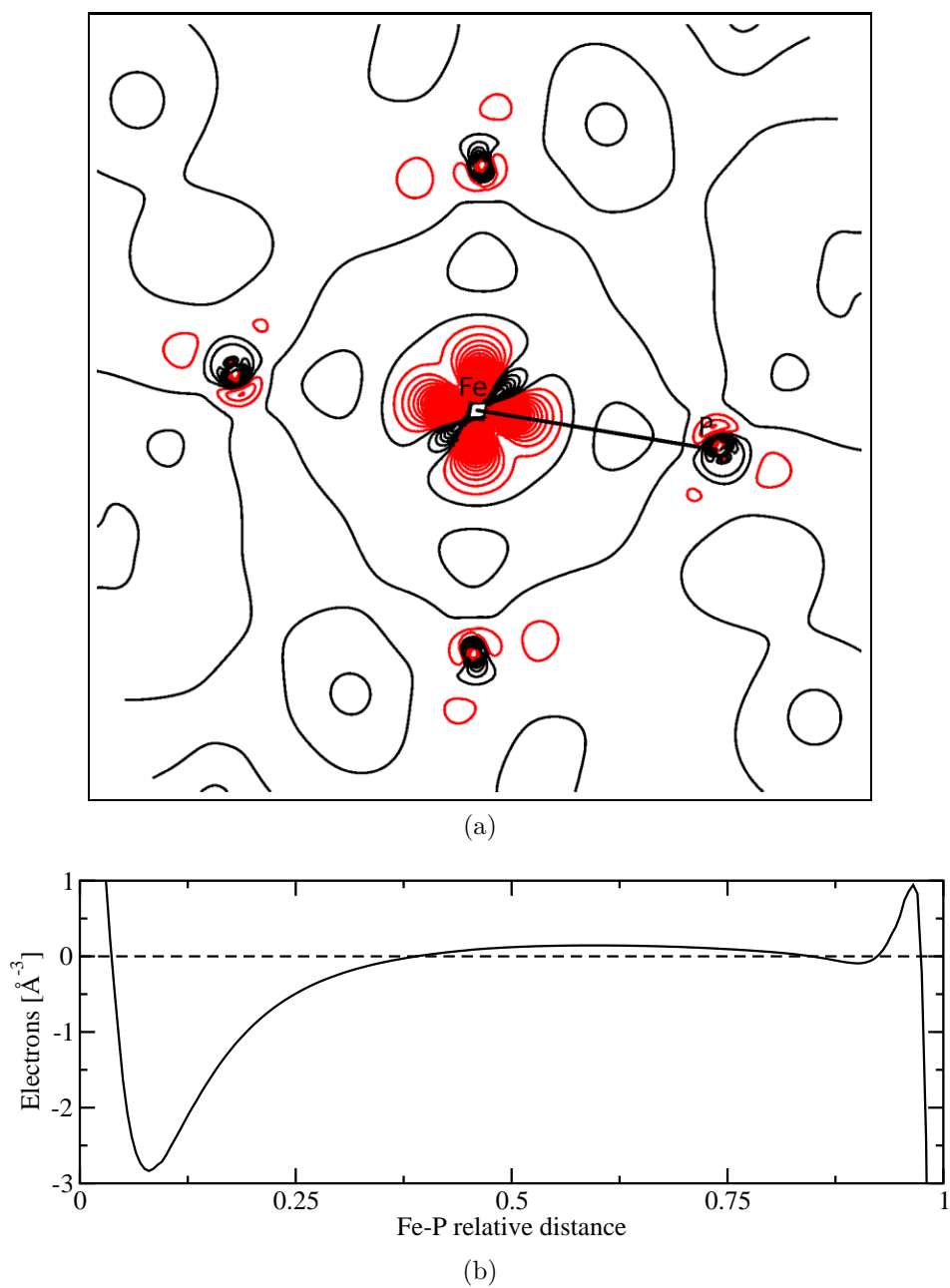}} \label{fig:countocta}}
\vspace{3mm}
\subfigure[]{\includegraphics[width=0.8\textwidth]{figures/figure3b.eps} \label{fig:plotocta_lafe4p12}}
\caption{(Color online) (a) Contour plot of the charge
density difference $\rho_b$ for LaFe$_4$P$_{12}$ in the octahedra cutout
(green/darkest plane in \figref{skutt}). Positive charge transfer
(black/dark) are drawn from 0 to 1.0 electrons/{\AA}$^{3}$, while the negative
(red/light) are drawn from -3 to 0 electrons/{\AA}$^{3}$. The contour spacing is
0.1 electrons/{\AA}$^{3}$. (b) Inter-ionic extraction of $\rho_b$ between Fe and P
(indicated by the black line in (a)). The distance is normalized.
} 
\label{fig:octa_lafe4p12}
\end{figure}

\begin{figure}[ht]
\centering
\subfigure[]{\fbox{\includegraphics[width=0.65\textwidth]{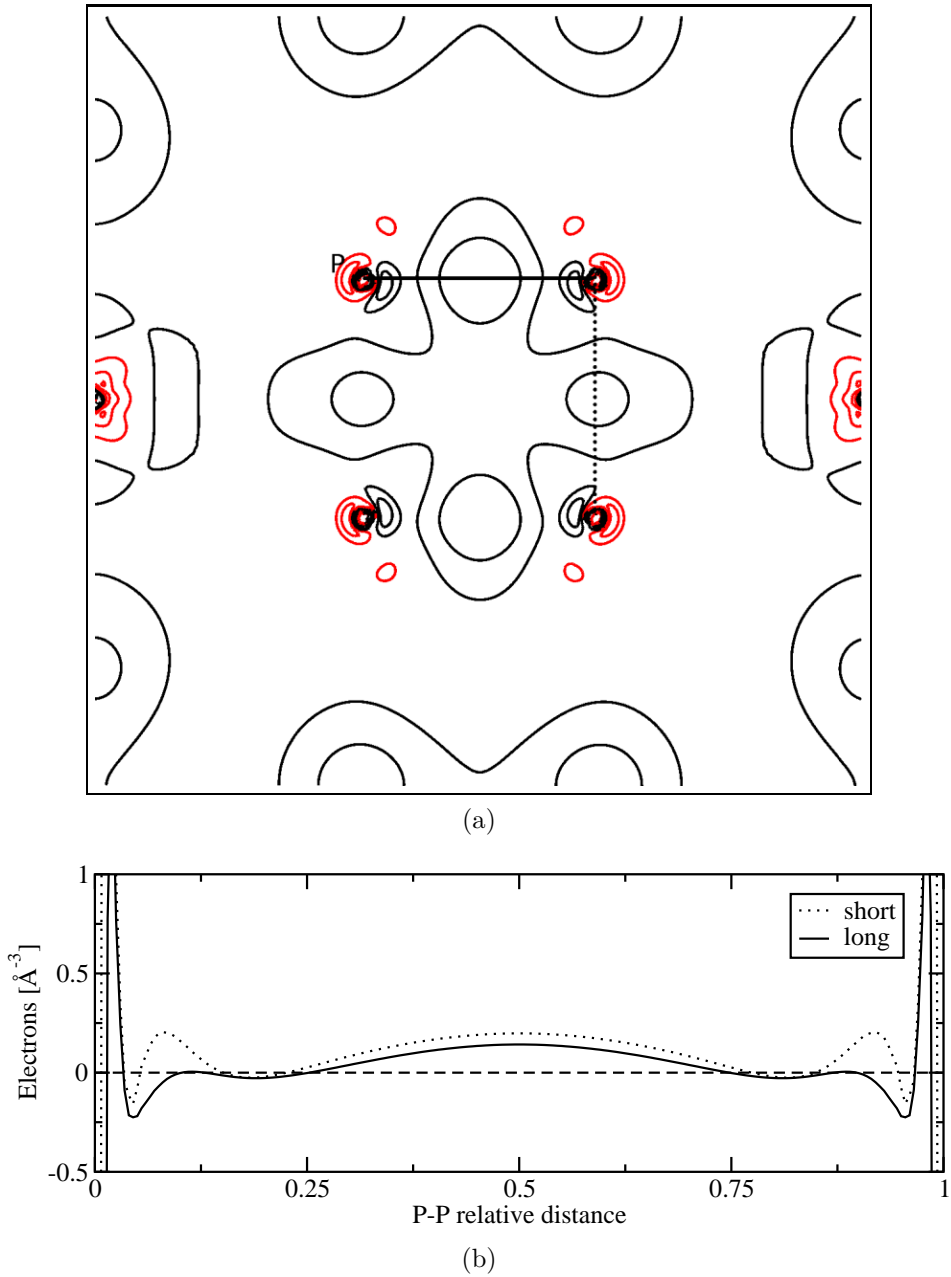}} \label{fig:countpnic}}
\vspace{3mm}
\subfigure[]{\includegraphics[width=0.8\textwidth]{figures/figure4b.eps} \label{fig:plotpnic_lafe4p12}}
\caption{(Color online) (a) Contour plot of the charge density difference $\rho_b$ for LaFe$_4$P$_{12}$ in the plane through the
P$_4$ ring (blue/brightest in \figref{skutt} with a shifted unit cell).
Positive charge transfer (black/dark) are drawn from 0 to
2.0 electrons/{\AA}$^{3}$, while the negative (red/light) are drawn from -0.5 to 0 electrons/{\AA}$^{3}$, both with a contour spacing of 0.1 electrons/{\AA}$^{3}$. (b)
Inter-ionic extraction of $\rho_b$ between two sets of P ions (indicated by the black lines in (a)). The straight and dotted
lines represent the short and long inter-ionic charge transfer
extractions P-P in the P$_4$ ring. These lines are indicated in
(a). The distances are normalized.
} 
\label{fig:pnic_lafe4p12}
\end{figure}

\begin{figure}[ht]
\centering
\subfigure[]{\fbox{\includegraphics[width=0.65\textwidth]{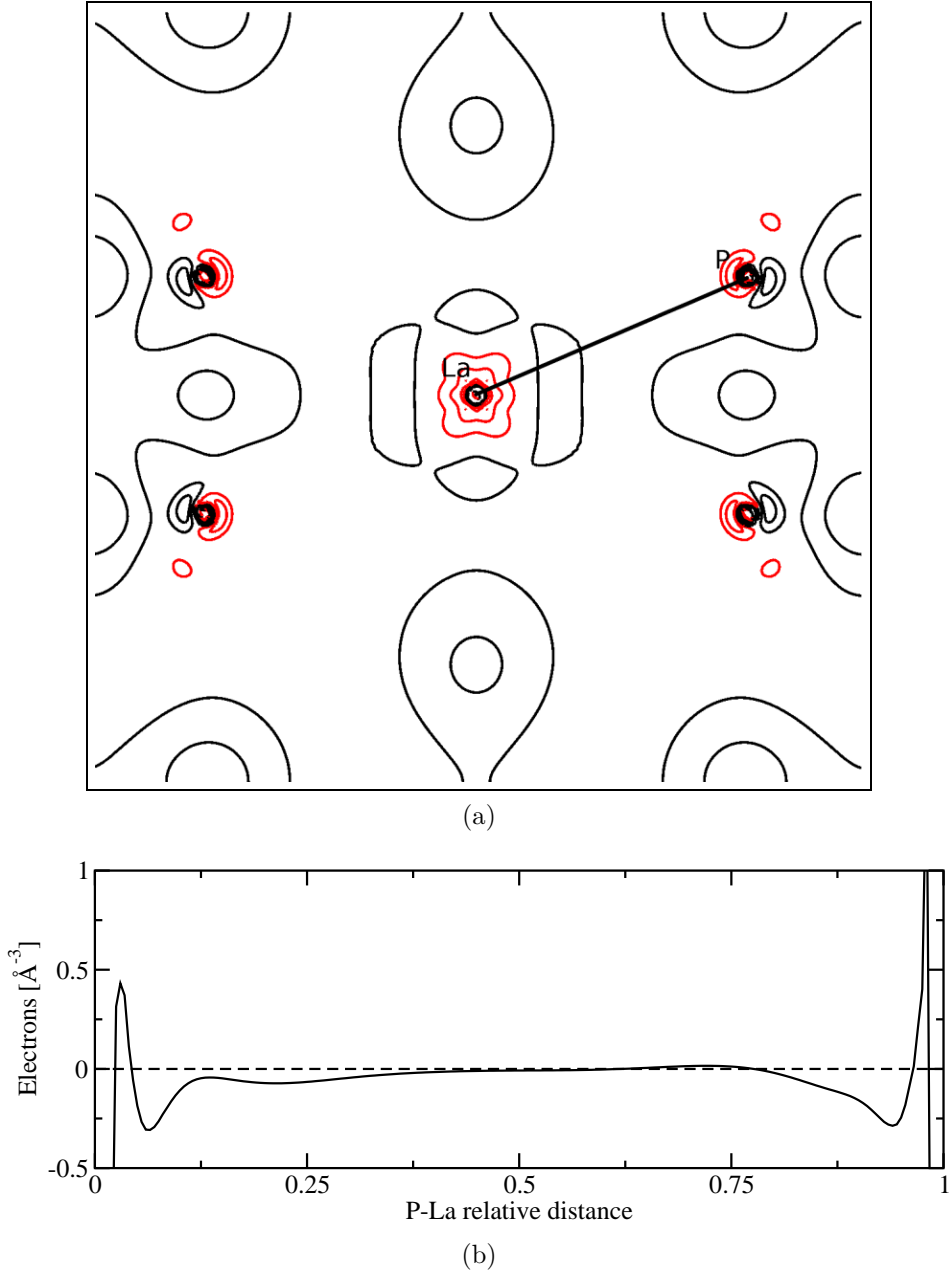}} \label{fig:countla}}
\vspace{3mm}
\subfigure[]{\includegraphics[width=0.8\textwidth]{figures/figure5b.eps} \label{fig:plotla_lafe4p12}}
\caption{(Color online) (a) Contour plot of the charge density difference $\rho_b$ for LaFe$_4$P$_{12}$ in the plane containing one
La surrounded by four P (blue/brightest plane in \figref{skutt}).
Positive charge transfer (black/dark) are drawn from 0 to
1.0 electrons/{\AA}$^{3}$, while the negative (red/light) are drawn from -0.5 to 0 electrons/{\AA}$^{3}$, both with a contour spacing of 0.1 electrons/{\AA}$^{3}$. (b)
Inter-ionic extraction of $\rho_b$ between P and La (indicated by
the black line in (a)). The distance is normalized.
} 
\label{fig:la_lafe4p12}
\end{figure}

Charge density differences $\rho_b$ have been calculated for
LaFe$_4$P$_{12}$ using the procrystal as a reference, $\rho_r = \rho_p$.
Contour plots of $\rho_b$ in the octahedral plane, the plane
containing the P$_4$ ring (020 in-plane shifted) and the
plane containing one La ion and four P ions (020) are
shown in \figref{octa_lafe4p12}, \figref{pnic_lafe4p12} and \figref{la_lafe4p12} respectively. They correspond
to the planes illustrated in \figref{skutt}. A line is indicated in
each contour plane which defines a line extraction of the
charge transfer in order to complement the contour plot.
The maximum along these lines, their relative distance
and the center of electronegativity are listed in \tabref{chargemax}.

Due to the lack of core electrons, there is a clear divergent behaviour close to each ion core, but this should
not be of importance to the charge transfer analysis presented. The distances have been normalized to a relative
distance $x$ in order to facilitate direct comparison between
the different compounds.

First of all, a significant charge depletion locally
around the Fe ion is clearly observed in \figref{octa_lafe4p12}. The charge
depletion is directional and facing the P ions, while depletion around the P ions is moderate. Also, a charge
buildup between the Fe and P ion is observed in \figref{octa_lafe4p12}
with a maximum value and a relative distance listed in
\tabref{chargemax}.

Going to the P$_4$ ring in \figref{pnic_lafe4p12}, parts of the charge transfer to the Fe-P binding can be seen as four negative lobes
around the P ions pointing out of the P$_4$ ring. These also
contribute to the positive P-P charge buildup. It is interesting to note that the charge transfer of the short
P-P bond is significantly different from its longer counterpart,
demonstrating that the shortest bond is the strongest.
As expected the maximum charge buildup is located in
the middle of both the short and long P-P bonds.

The P-La charge transfer is given in \figref{la_lafe4p12}. Compared
to the Fe-P bond, a maximum value is observed closer
to the La ion. In addition significant charge movement
close to the P and La ion is confirmed. The depletion
close to the P ion is similar to the depletion around P for
the long P-P bond, which is facing La.

\begin{figure}[ht]
  \centering
\subfigure[]{\fbox{\includegraphics[width=0.65\textwidth]{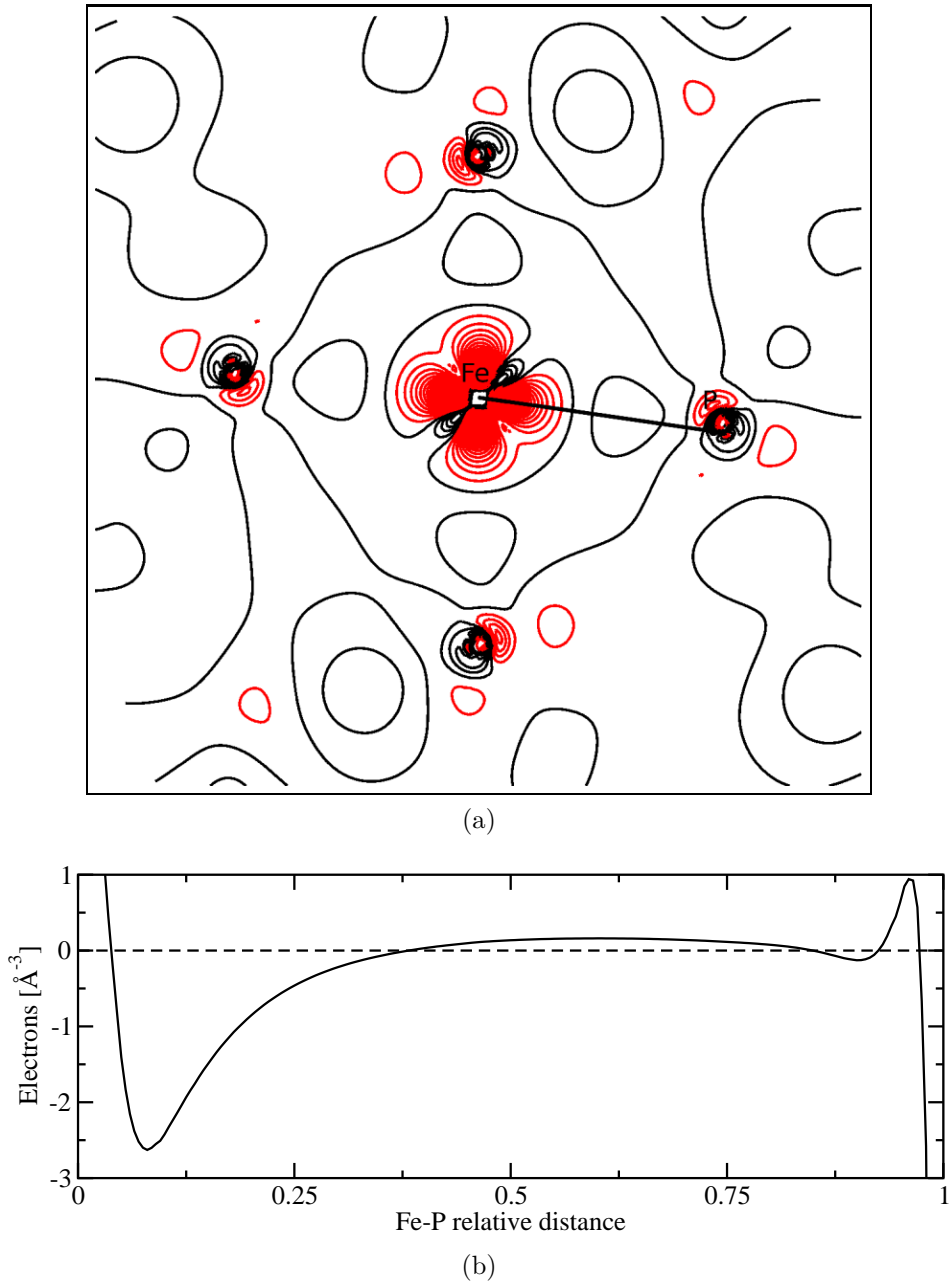}} \label{fig:octa_fep3_plane}}
\vspace{3mm}
\subfigure[]{\includegraphics[width=0.8\textwidth]{figures/figure6b.eps} \label{fig:octa_fep3_line}}
\caption{(Color online) (a) Contour plot of the charge density difference $\rho_b$ for FeP$_3$ in the octahedra (green/darkest
plane in \figref{skutt}). Positive charge transfer (black/dark) are
drawn from 0 to 1.0 electrons/{\AA}$^{3}$, while the negative (red/light) are drawn from -3 to 0 electrons/{\AA}$^{3}$, both with a contour spacing of
0.1 electrons/{\AA}$^{3}$. (b) Inter-ionic charge extraction of $\rho_b$ between Fe
and P (indicated by the black line in (a)). The distance is normalized.
} 
\label{fig:octa_fep3}
\end{figure}

An important question to raise is how much the La ion
contributes to the Fe-P charge transfer in LaFe$_4$P$_{12}$ compared to the pure Co to Fe substitution going from CoP$_3$
to FeP$_3$. To illustrate this we have calculated the hypothetical FeP$_3$ structure based on the atomic positions
and unit cell parameters of CoP$_3$. The FeP$_3$ structure
was not relaxed. In \figref{octa_fep3} we give the contour plot of the
octahedral plane of FeP$_3$ and the inter-ionic Fe-P line extraction. Comparing \figref{octa_lafe4p12}(a) and \ref{fig:octa_fep3}(a) shows that the
overall differences between FeP$_3$ and LaFe$_4$P$_{12}$ in the octahedral plane are small close to the the Fe ion, while
clear differences are observed in the vicinity of the P ion.

\begin{figure}[ht]
\centering
\includegraphics[width=0.8\textwidth]{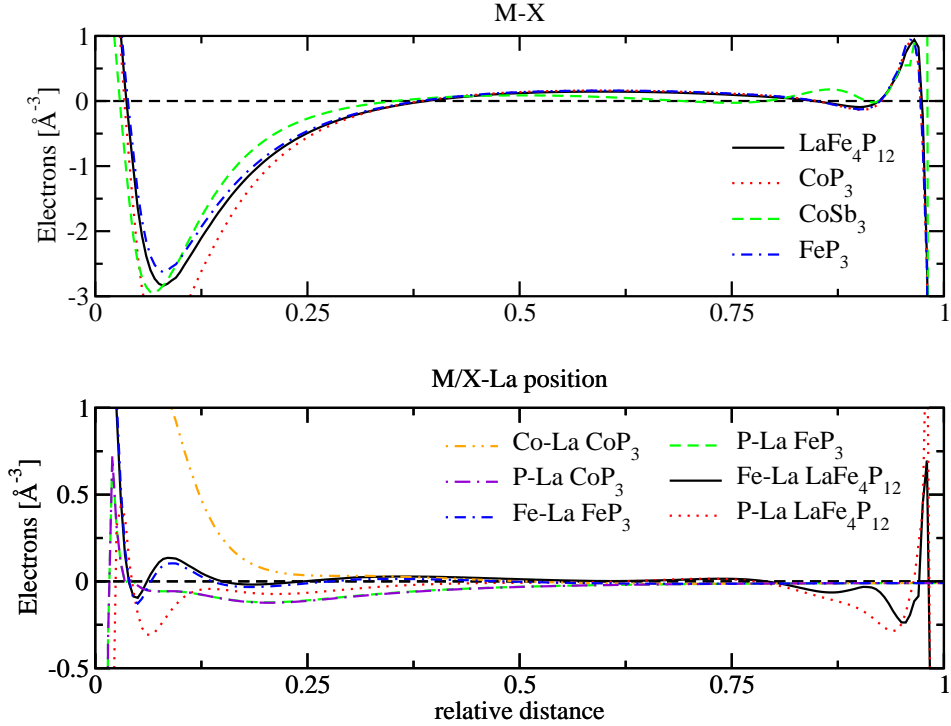}
\caption{(Color online) Inter-ionic extraction of the charge
density difference $\rho_b$ between $M$-$X$ (upper) and $M$/$X$-La(lower). The transition metal Co, Sb and the pnictogens
P, Sb are designated $M$ and $X$, respectively. In the upper
a comparison between LaFe$_4$P$_{12}$ (solid line), CoP$_3$ (dotted
line), CoSb$_3$ (dashed line) and FeP$_3$ (dashed-dotted line) is
given, while in the lower a comparison between Fe-La (solid
line) and P-La (dotted line) in LaFe$_4$P$_{12}$, Co-''La'' (short
dashed-double-dotted line) and P-''La'' (long-dashed-dotted line) in
CoP$_3$, Fe-''La'' (dashed-dotted line) and P-''La'' (dashed) in
FeP$_3$ is given. ``La'' represents Wyckoff position a where e.g. La
atoms would have been located. The distances are normalized.
}
\label{fig:tm_and_la}
\end{figure}

\begin{figure}[ht]
\centering
\includegraphics[width=0.8\textwidth]{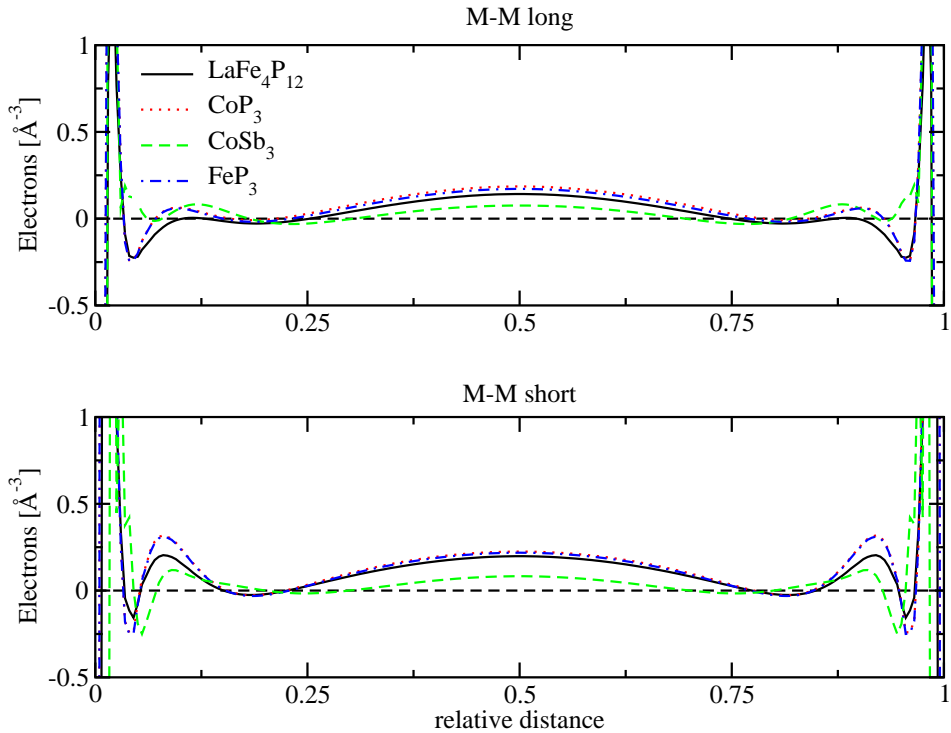}
\caption{(Color online) Inter-ionic extraction of the charge
density difference $\rho_b$ between the short $X$-$X$ bonds (lower)
and its longer counterpart (upper). The pnictogens P and Sb
are designated $X$. A comparison between LaFe$_4$P$_{12}$ (solid
line), CoP$_3$ (dotted line), CoSb$_3$ (dashed line) and FeP$_3$
(dashed-dotted line) is given. The distances are normalized.
}
\label{fig:pnic_both}
\end{figure}

The complete set of relevant charge transfers $\rho_b$ are
compared for CoP$_3$, CoSb$_3$, FeP$_3$ and LaFe$_4$P$_{12}$ in \figref{tm_and_la}
and \ref{fig:pnic_both}. Maximum values and their relative distances are
listed in \tabref{chargemax}. In the upper part of \figref{tm_and_la},
a charge buildup maxima for CoP$_3$ and FeP$_3$ is observed, slightly
shifted towards the P ions. Differences between CoP$_3$
and FeP$_3$ close to the $M$ ion are also observed, while the
differences closer to the P ions are marginal. Close to
the $M$ ions the largest charge depletion is found in CoP$_3$
followed by CoSb$_3$, LaFe$_4$P$_{12}$ and FeP$_3$. The location of
the smallest charge buildup maximum for Co-Sb is in the
middle of the $M$-$X$ bond. In addition a second larger
maximum closer to the P ion is observed. This extra
maximum is completely lacking for the other compounds.

In the lower part of \figref{tm_and_la}, both P-La and $M$-La inter-ionic charge transfers are illustrated for CoP$_3$, FeP$_3$ and
LaFe$_4$P$_{12}$. For CoP$_3$ and FeP$_3$ the virtual position of a
filler atom (Wyckoff position a) is used for direct comparison of the
electron density around this position. The
Fe-La charge transfer in LaFe$_4$P$_{12}$ has three well defined
charge difference maxima. The two smallest are about an
order of magnitude smaller than the values for the $M$-$X$
and $X$-$X$ bonds. The inter-ionic charge transfer extraction between Fe and the virtual “La” position for FeP$_3$
follows that of LaFe$_4$P$_{12}$ close to the Fe ion; the largest
difference between the two compounds (except the obvious difference close to the La position) is observed in
the first peak close to the Fe ion. When La is added, a
change in charge depletion close to the P ion in the P-La
extraction is observed. However, this depletion is larger
for LaFe$_4$P$_{12}$ than FeP$_3$, thus signifying a redistribution of charge close the P ion when La is added.
A noteworthy observation is the difference between Fe
and Co for the $M$-La bond direction. Close to the Co
ion there is a significant charge buildup, which is not so
evident close to the Fe ion.

From \figref{pnic_both} it is
clear that the P-P charge transfer changes upon addition of
La. A reconfiguration of both the long and
short bonds is observed. The charge buildup at the midpoint and close
to P decreases when La
is added. 
For the P-P bond in CoP$_3$ the short to long charge
buildup maximum ratio is 0.83. For FeP$_3$ the ratio is
0.79, while it changes to 0.72 for LaFe$_{4}$P$_{12}$. Also,
when La is added there is a reconfiguration; the short
and long P-P distances are exchanged compared to FeP$_3$. 
When turning to the antimonide, the Sb-Sb bond exhibit smaller charge
buildup relative to the P-P bond with a short to long charge
buildup maximum ratio of 0.91.
\begin{table}[ht]
\caption{
Charge density difference maxima $\rho_{max}$ and relative distance
$x$ for all significant bond directions for CoP$_3$, FeP$_3$ and
LaFe$_4$P$_{12}$. In addition the center of Pauling electronegativity
$\chi^∗$ defined in the text is given (electronegativities are listed in \tabref{chargemax}).}
\begin{center}
\begin{tabular*}{0.8\textwidth}{@{\extracolsep{\fill}}lllll}
\br
&& $\rho_{b}^{max}$\mbox{[electrons/{\AA}$^{3}$]}& $x$ &$\chi^{*}$\\
\mr
CoP$_3$ &  & & &\\
&Co-P &0.163 & 0.6 &0.54\\
&P-P  &0.188 & 0.5 &0.5\\
&     &0.227 & 0.5 &\\ 
FeP$_3$ &  & & &\\
&Fe-P &0.160 & 0.6 &0.55\\
&P-P  &0.172 & 0.5 &0.5\\
&     &0.218 & 0.5 &0.5\\ 
CoSb$_3$&  & & &\\  
&Co-Sb &0.086 & 0.5 &0.52\\
&     &0.117 & 0.87&0.52\\
&Sb-Sb  &0.076 & 0.5 &0.5\\
&     &0.083 & 0.5 &0.5\\
LaFe$_4$P$_{12}$ &  & & &\\
&Fe-P &0.144 & 0.59&0.55\\
&P-P  &0.142 & 0.5 &0.5\\
&     &0.198 & 0.5 &0.5\\
&Fe-La&0.167 & 0.09&0.38\\
&     &0.029 & 0.36&0.38\\
&     &0.015 & 0.74&0.38\\
&P-La &0.016 & 0.72&0.33\\
\br
\end{tabular*}
\end{center}
\label{tab:chargemax}
\end{table}
\subsection{Quantitative charge analysis}\label{sec:quantitative}
\subsubsection{Charge depletion}\label{sec:chargedep}
In \tabref{spherint} the spherical charge depletion number $\Delta n_d$
as defined in \equref{diffchange} and its mean value $\langle \Delta n_d \rangle$ defined
in \equref{dipoleeq} are given. From this table we observe that the
charge depletion number around the $M$ ion is largest in
CoP$_3$ followed by LaFe$_4$P$_{12}$, FeP$_3$ and CoSb$_3$. The
analogues sequence for the $X$ ions is CoSb$_3$, LaFe$_4$P$_{12}$, CoP$_3$ and FeP$_3$. Similar values of $\langle \Delta n_d \rangle$ for CoP$_3$, FeP$_3$
and LaFe$_4$P$_{12}$ are observed, while it is less for Co and
larger for Sb in CoSb$_3$ compared to the other compounds.

The charge depletion around the La ion is largest of all
ions, but its mean value is similar to the ones around Co
and Fe. The ratio of the depletion number between Co
and Fe is almost the same as the relative valence charge
difference between the two atoms. It is also observed
that the addition of La increases the Fe and P depletion.
The depletion cutoff $r_d$ is comparable for the P ions between CoP$_3$, CoSb$_3$ and FeP$_3$, while it increases slightly
for LaFe$_4$P$_{12}$. Comparable values are also observed among
the Co ions and Fe ions in the respective structures.
\begin{table}[ht]
\caption{Spherical depletion number $\Delta n_d$ and their mean value
$\langle \Delta n_d \rangle$, the cutoff radii $r_d$ and the Pauling
electronegativities\cite{exp:allred,exp:pauling}$\chi$ are shown. The
covalent radii $r_c$\cite{covalent} are also listed for comparison.}
\begin{center}
\begin{tabular*}{0.8\textwidth}{@{\extracolsep{\fill}}lllllll}
\br
&& $\Delta n_{d}$\mbox{[electrons]} & $\langle \Delta n_{d}\rangle
$\mbox{[electrons]}&$r_{d}$\mbox{[{\AA}]} &$r_{c}$\mbox{[{\AA}]}& $\chi$\\
\mr
CoP$_3$ &&&&&\\
&Co &-0.241&-0.099& 0.804& 1.260&1.88\\
&P  &-0.048&-0.017& 0.803& 1.070&2.19\\
FeP$_3$ &&&&&\\
&Fe &-0.218&-0.092& 0.841& 1.320&1.83\\
&P  &-0.047&-0.016& 0.803& 1.070&2.19\\ 
CoSb$_3$&&&&&\\  
&Co &-0.161&-0.067& 0.803& 1.260&1.88\\
&Sb &-0.110&-0.046& 1.150& 1.390&2.05\\
LaFe$_4$P$_{12}$ &&&&&\\
&Fe &-0.221&-0.094& 0.841& 1.320&1.83\\
&P  &-0.056&-0.020& 0.838& 1.070&2.19\\
&La &-0.242&-0.108& 1.986& 2.070&1.10\\
\br
\end{tabular*}
\end{center}
\label{tab:spherint}
\end{table}
\subsubsection{Bader analysis}\label{sec:bader}
While the intuitive spherical integration is bound to
fail in systems with complicated charge geometries, the
Bader analysis is one of the few techniques with a well defined charge
partitioning\cite{theo:bader}. Despite the clear definition, it may be difficult to compare different compounds
and their relative charge transfer, due to the variation
of the Bader volumes. If the volumes are too different,
\begin{table}[ht]
\caption{Bader analysis of CoP$_3$ , FeP$_3$ , CoSb$_3$ and
LaFe$_4$P$_{12}$ . Both the charge and the minimum distance to
the Bader surface are given. For the pnictogens all four non-
equivalent (in the Bader analysis) values are listed. The
charge difference is relative to the formal valence charge of
the free atom.
}
\begin{center}
\begin{tabular*}{0.8\textwidth}{@{\extracolsep{\fill}}llllrr}
\br
&& \multicolumn{3}{c}{valence electrons} & Min. distance[\AA]\\
&& Atom & Crystal & Difference &\\
\mr
CoP$_3$ &&&&&\\
&Co & 9.0 &8.89 & -0.11 &1.02 \\
&P & 5.0 & 5.01& 0.01&1.11 \\
& & & 5.04 & 0.04 & 1.11 \\
& & & 5.04 & 0.04 & 1.13 \\
& & & 5.06 & 0.06 & 1.13\\
FeP$_3$ &&&&&\\
&Fe & 8.0 &7.78 & -0.22 &1.00 \\
&P & 5.0 & 5.02& 0.02& 1.09 \\
& & & 5.07 & 0.07 & 1.09\\
& & & 5.08 & 0.08 & 1.12\\
& & & 5.13 & 0.13 & 1.12 \\
CoSb$_3$&&&&&\\  
&Co & 9.0& 9.32& 0.32&1.13\\
& Sb& 5.0& 4.88 & -0.12 & 1.34\\
& & & 4.89 & -0.11 & 1.34 \\
& & & 4.90 & -0.10 & 1.34 \\
& & & 4.91 & -0.09 & 1.34 \\
LaFe$_4$P$_{12}$ &&&&&\\
&Fe& 8.0& 7.81& -0.19 &1.03\\
&P &5.0 & 5.17 & 0.17 & 1.13\\
& & & 5.19 & 0.19 & 1.13\\
& & & 5.19 & 0.10 & 1.15\\
& & & 5.22 & 0.22 & 1.15\\
&La & 11.0 &9.47 & -1.53&1.52 \\
\br
\end{tabular*}
\end{center}
\label{tab:baderanal}
\end{table}
a comparative charge transfer analysis may be quantitatively unreliable. It is difficult to tell whether the different volumes are due to artifacts of the analysis or different nature of the bonds. Nevertheless, the analysis is
well defined and has previously showed to give reliable
charges and good comparison conditions even for different volumes.
Bader analysis data for CoP$_3$, CoSb$_3$, FeP$_3$ and
LaFe$_4$P$_{12}$ are given in \tabref{baderanal}. 

For the $X$ ions four different values are listed corresponding to nonequivalent positions (in the Bader analysis). Even though
the differences are small the largest ones are outside estimated error margins. The picture of the charge transfers
drawn in the previous sections is confirmed. However,
the Bader analysis fails to determine the expected transfer in CoSb$_3$, in contrast to previous work\cite{theo:ghosez}. Also,
the variation among the antimonides is not present for
CoSb$_3$. This could indicate a failure of the code used in
this work or the Bader analysis itself. The volume filling of the Bader analysis is rather good.
The errors for the charge filling of the unit cell is below
one percent for most compounds if the average $X$
values are used. However, significant differences are observed confirming the ambiguity of the code used in this
work, the core grids or the Bader analysis itself.
\subsection{Discussion}\label{sec:discussion}
We will first discuss the character of different bonds.
The center of electronegativity $\chi^*$ suggests that a charge
maximum should occur at 0.54, 0.52 and 0.52 relative distance for the Co-P, Fe-P and Co-Sb bonds if the bonds
are purely covalent. From \tabref{chargemax} this is not exactly the
case for Co-P and Fe-P bonds, but close enough to conclude that they have a co-ordinated covalent character, in
agreement with earlier suggestions\cite{exp:uher}. The small deviations are probably attributed to the hybridization set up
in the $M$P$_6$ octahedra\cite{exp:uher}. It is interesting to note that
the Co-Sb bond seems to exhibit a stronger covalency than the other $M$-$X$ bonds, in agreement with previous
work\cite{exp:grosvenor} and the electronegativity differences. The P-P
bonds are obviously covalent in character illustrated by
the charge buildup between the ions. It is more complicated to
conclude on a Fe-La bond; however, signs of a covalent character is present in this study. The P-La
bond does not seem to possess a covalent character, in
agreement with earlier suggestions\cite{exp:grosvenor,exp:uher}.
The charge depletion around each ion seems to follow
the general trends predicted by simple electronegativity
considerations. 

The depletion is largest for Co in CoP$_3$, the structure with ions exhibiting the largest difference
in electronegativities (of the unfilled structures). It could
be expected from the electronegativities that Fe is more
depleted in FeP$_3$ than Co in CoP$_3$. But since the Fe atom
contains one less electron the relative depletion number is
smaller. 

In CoSb$_3$, the difference in electronegativity between $M$ and $X$ is smaller than for CoP$_3$ and FeP$_3$. This
leads to a decrease of charge depletion around the Co ion
and an increase around the Sb ion. From \figref{tm_and_la} an extra
charge maximum can be observed close to Sb ion, which is
completely lacking in the other compounds. This is probably attributed to the d-electrons in the outer core shell
of Sb, which implies a different kind of bond scheme than
for P. The peak close to the Sb ion contradicts the previous
electronegativity agreements. 

The small difference
between CoP$_3$ and FeP$_3$ confirms that the main differences between
the two compounds are found in the vicinity of
$M$. 

Iso-surface studies (not shown) also confirm the suggested sp$^3$-like hybridization\cite{exp:uher} of P, where the sp$^{3}$-like
lobes are facing P and Fe. For the $X$-$X$ bonds a greater
charge buildup between P than between Sb is observed.

Adding La to FeP$_3$ increases Fe and P charge depletion, indicating a reorganization of charge to set up of
Fe-La and P-La bonds, stabilizing LaFe$_4$P$_{12}$ compared
to FeP$_3$. This is also reflected in the differences north
and south of the P positions in figure \ref{fig:octa_fep3} and \ref{fig:octa_lafe4p12} for FeP$_3$
and LaFe$_4$P$_{12}$ respectively. A relatively weak binding
of La to neighbouring P ions is observed. The maximum
charge buildup between La and P is significantly less than
what was found in the Fe-P bond (about a tenth). Close to the P ions
the setup of a La-P bond and the modification of the P-P
bond imply changes in the charge, which is observed.
The difference between the $X_{4}$ ring ratio of
CoP$_3$ and CoSb$_3$ is attributed to the increased size of
the Sb atoms which will have greater influence on the
short bonds. 

The results from the Bader analysis generally follow
simple electronegativity considerations. This also holds
true when La is added to the FeP$_3$ system. For LaFe$_4$P$_{12}$
the Bader analysis indicates a rather large increase of
charge at the P positions, while the charge difference of
Fe is similar between FeP$_3$ and LaFe$_4$P$_{12}$. This is in
agreement with the results from the depletion numbers. The different
values for the $X$ ions are related to the finite grid
resolution.
The failure to reproduce physical results for CoSb$_3$ (the
charge transfer behaviour is expected to follow that of
CoP$_3$) is either attributed to the program used in this
work, the core grid of Sb or the Bader analysis itself. However, it
should be emphasized that within reasonable grid resolutions this
behaviour does not change. The core grids and the Bader analysis are not used during the
calculation of depletion numbers where such a discrepancy for CoSb$_3$ is lacking.

Furthermore, the $P_{4}$ ring ratio changes quite drastically from CoP$_3$ to
LaFe$_4$P$_{12}$. In CoP$_3$ the shortest length is facing the vacant filler position, while in LaFe$_4$P$_{12}$ this is opposite.
Also, the rectangular P$_4$ ring in CoP$_3$ is transformed into
a more quadratic shape in LaFe$_4$P$_{12}$ in correspondence
with what happens when CoSb$_3$ is filled with
Ce\cite{exp:kitagawa}. The La atoms want to bond to the nearby dodecahedra of P ions by
redistributing charge from the short P bond to the bond between P and La. This results in a weaker effective P-P bonds
close to the La ion, thus increasing the distance between the
P ions. As a consequence there is an interchange of
the short and long P-P bonds between CoP$_3$ and
LaFe$_4$P$_{12}$.

Adding to the above analysis, differences between $M$-P
bonds in CoP$_3$ and LaFe$_4$P$_{12}$ are small. There is virtually
no difference in the charge buildup except close to the
$M$ ion. Thus, the depletion around the Fe ion does not
contribute directly in a modified Fe-P binding when La
is added, a further evidence of new Fe-La bonds. Due
to the similarities close to the Fe ions between FeP$_3$ and
LaFe$_4$P$_{12}$ it is clear that the establishment of P-La bonds
would primarily modify the P$_4$ ring.

From \figref{la_lafe4p12} elongated charge buildups between the
La ion and the long P-P bond is observed. These are aligned along the crystal axes. The
bonding between P and La is likely not directional between the ions such
that La interacts with the closest P-P bonds by establishing shared states with
both P ions. The large depletion cutoff $r_d$ suggest an extended
depletion around La. Yet, the elongated bonds are established in this
zone, further strengthening the evidence that the La-P bonds are
non-directional and elongated. This is highly relevant to the
``rattling'' behaviour of the filled skutterudites. Vibrations of La
along the crystal axes would imply a response from the P dodecahedra and hence be correlated throughout the structure. On the
other hand, vibrations along directions between the elongated bond
(e.~g. in the north-east direction from La in \figref{la_lafe4p12})
are possibly more uncorrelated and anharmonic. Both pictures are supported in previous
studies\cite{exp:keppensratlers,exp:saleskeppens,exp:chris,exp:koza}
which apparently are contradictory. But
we believe that the combination of both pictures is vital for
understanding the reduced thermal conductivity. A possible explanation of the phonon dampening
in skutterudites is as follows. Take as an example a phonon wave that
involves movement of the P$_{4}$ ring parallel to the
crystal axes. This movement would be transferred onto the
elongated bond, but as lanthanum moves towards the next elongated bond
it is deflected towards to openings due to the edged shape of the
elongated bond (e.~g. moving in the almost north direction from La in
\figref{la_lafe4p12}). As a consequence the phonon wave vector is
changed and the phonon wave is damped.
\section{Conclusions}\label{sec:conclude}
In this work we have investigated the charge transfer in
the CoP$_3$, CoSb$_3$, LaFe$_4$P$_{12}$ and the hypothetical FeP$_3$
skutterudite compounds by using the corresponding procrystals as
charge references. 

It was demonstrated that the
alternative references, like the standard state reference,
may lead to inconsistent results.

General agreements with simple electronegativity
considerations were shown. The covalent character of the $M$-$X$ and $X$-$X$ bonds
was confirmed, while a slight ionic character was detected for the
P-La bond. It was also shown that the addition of La results in
a redistribution of charge within the P$_4$ rings.

A unique bonding scheme between P and La was proposed. We showed that elongated bonds
close to the La ion were established when La was added. These
elongated bonds were aligned along the crystal axes facing the P$_{4}$
ring. Vibrations of the La
ion along the crystal axes would then be closely connected with the
surrounding P$_{4}$ rings. In contrast, vibrations of La along directions
between the elongated bonds would be more uncorrelated throughout the structure. The
combination yields a quasi-correlated
motion of the La ion, which supports recent\cite{exp:koza,exp:chris} and
previous\cite{exp:keppensratlers,exp:saleskeppens} work. We proposed a
possible direct explanation of the phonon dampening in the
skutterudites being caused by a change of the phonon wave vector. This change comes as a result of the elongated
bonds. Studies are in progress to
investigated the spatially resolved movement of La ions to further enlighten
this picture. 

\ack{
The authors would like to acknowledge support from the Norwegian
Research Council and the NOTUR project. In addition we would like to thank
Simone Casalo, Ragnhild S{\ae}terli, Ole Bj{\o}rn Karlsen and Jon
Nilsen for fruitful discussions and ideas.}

\bibliography{reffs}
\end{document}